# Parallel Polarization State Generation

Alan She[1*] and Federico Capasso[1*]

[1]Harvard John A. Paulson School of Engineering and Applied Sciences, Cambridge, MA 02138

*To whom correspondence should be addressed;
E-mail: ashe@seas.harvard.edu; capasso@seas.harvard.edu.

   The control of polarization, an essential property of light, is of wide scientific and technological interest[1,2]. The general problem of generating arbitrary time-varying states of polarization (SOP) has always been mathematically formulated by a series of linear transformations, i.e. a product of matrices, imposing a serial architecture. Here we show a parallel architecture described by a sum of matrices. The theory is experimentally demonstrated by modulating spatially-separated polarization components of a laser using a digital micromirror device that are subsequently beam combined. This method greatly expands the parameter space for engineering devices that control polarization. Consequently, performance characteristics, such as speed, stability, and spectral range, are entirely dictated by the technologies of optical intensity modulation, including absorption, reflection, emission, and scattering. This opens up important prospects for polarization state generation (PSG) with unique performance characteristics and applications in spectroscopic ellipsometry, spectropolarimetry, communications, imaging, and security.

   In everyday use, SOPs are commonly met in the so-called "degenerate polarizations" as linearly and circularly polarized light but are in general elliptically polarized[3,4]. To describe and control the polarization of light, the projections of the electric field onto an orthogonal bases and their relative phase relation must be known and are mathematically represented by the Jones vector and Stokes Parameters[1,5] (see Supplementary Information).

   In conventional serial architectures, the polarization of an input beam, $E_{in}$, may be transformed into any arbitrary output polarization, $E_{out}$, through a product of Jones matrices $M_n$ corresponding to variable optical elements, each of which has a degree of freedom, $\rho_n$: $E_{out} = M_N(\rho_N)...M_2(\rho_2)M_1(\rho_1)E_{in}$. Commonly found implementations of serial PSGs use optical elements that introduce suitable phase shifts or birefringence, which are represented by a product of at least two Jones matrices. These include devices such as rotating waveplates[6], Babinet-Soleil compensators[1], Berek rotary compensators[7], fiber coil polarization controllers[8], Faraday rotators[9], fiber squeezers[10], polarization Michelson interferometers[11], degree of polarization generators[12], lithium niobate electro-optics[13]; liquid crystals[14]; and on-chip photonic circuits[15,16]. Figures of merit that characterize the performance of these devices include temporal response, stability, mechanical fatigue, insertion loss, SOP accuracy[17], and operating wavelength range.

   To develop a parallel architecture, we revisit the Fresnel-Arago interference laws, which state that light beams of orthogonal polarizations cannot interfere[18,19]. Beams that are coherent, however, create a linear superposition to produce a new SOP. In our approach, we propose PSG by combining a limited set of prepared SOPs, which we refer to here for convenience as the "Stokes Basis Vectors" (SBVs), and are not necessarily linearly independent in the conventional sense. *By modulating the intensities of a number of beams corresponding to a set of SBVs and combining them, we are able to generate any arbitrary output SOP (Fig. 1).*

   A set of SBVs labeled by $n$ can be described as follows as Jones vectors:

$$\mathbf{C}_n = \begin{pmatrix} C_{0nx} \\ C_{0ny}e^{i\theta_n} \end{pmatrix} e^{i\phi_n} = \begin{pmatrix} \tilde{C}_{nx} \\ \tilde{C}_{ny} \end{pmatrix} \tag{1}$$

where $C_{0nx}$ and $C_{0ny}$ are real coefficients, $\theta_n$ is the relative phase difference between polarization components, $\phi_n$ is the global phase, and $\tilde{C}_{nx}$ and $\tilde{C}_{ny}$ are complex amplitudes of the electric field. By linearly combining $N$ SBVs of equation (1) multiplied by modulation parameters, $\alpha_n$ (here real and positive scalar quantities corresponding to intensity modulations when squared), the resultant electric field can be expressed as the following:

$$\mathbf{E} = \alpha_1 \mathbf{C}_1 + \alpha_2 \mathbf{C}_2 + \cdots + \alpha_n \mathbf{C}_n \tag{2}$$

While the global phase of each SBV, $\phi_n$, does not affect its SOP, relative phase is an important factor in the interference between the SBVs, and its physical origin is the phase shift measured at the location where beams combine; $\phi_n$ can be tuned by changes in optical path length or by other means, such as resonant optical elements. It is shown later that the combination of a minimum of four SBVs is required to generate arbitrary SOPs, so that any desired Stokes vector can be mapped to four modulation parameters: $(S_1, S_2, S_3) \rightarrow (\alpha_1, \alpha_2, \alpha_3, \alpha_4)$. In the case of four SBVs, equation (2) can be rewritten as the following real matrix equation:

$$
\begin{pmatrix}
C_{1x}^{re} & C_{2x}^{re} & C_{3x}^{re} & C_{4x}^{re} \\
C_{1x}^{im} & C_{2x}^{im} & C_{3x}^{im} & C_{4x}^{im} \\
C_{1y}^{re} & C_{2y}^{re} & C_{3y}^{re} & C_{4y}^{re} \\
C_{1y}^{im} & C_{2y}^{im} & C_{3y}^{im} & C_{4y}^{im}
\end{pmatrix}
\begin{pmatrix}
\alpha_1 \\
\alpha_2 \\
\alpha_3 \\
\alpha_4
\end{pmatrix}
=
\begin{pmatrix}
E_x \cos\phi \\
E_x \sin\phi \\
E_y \cos(\theta + \phi) \\
E_y \sin(\theta + \phi)
\end{pmatrix}
\tag{3}
$$

where $\theta$ and $\phi$ are defined as in equation (1). This can be solved for real and positive $\alpha_n$ given a set of SBVs represented by the square matrix on the left hand side and the desired SOP given by the right hand side. The square values of the calculated $\alpha_n$ are used to modulate the intensities of the SBVs for final PSG. Additionally, the number of SBVs can be increased and each prepared with well-defined $\phi_n$ in order to add the capability of phase control to the generated SOP.

Polarization modulation can be visualized as dynamic polarization trajectories on the surface of the Poincaré sphere (Fig. 2a,b). For example, the linear combination of any two SOPs can be varied in order to create a line of SOPs on the Poincaré sphere: $\mathbf{E} = \alpha \mathbf{C}_1 + (1 - \alpha)\mathbf{C}_2$, in which two SOPs, $\mathbf{C}_1$ and $\mathbf{C}_2$ (that could be SBVs), are parameterized by $\alpha$ that is varied from 1 to 0 (Fig. 2a). Combining SOPs generates new SOPs by way of interference; depending on their relative phase, paths with varying curvature can be generated (Fig. S3). In order to deviate from this path, a third SOP, $\mathbf{C}_3$, must be introduced to provide one more degree of freedom, which expands the generable SOPs from a line to a surface (region). Within an arbitrary set of SBVs, each subset of three SBVs (C1, C2, and C3) can generate a surface bounded by the trajectories connecting each pair of SBVs (C1 and C2, C1 and C3, C2 and C3). Then arbitrary trajectories can be generated within this allowable surface, such as spiral or even chaotic trajectories (Fig. 2c,d and Supplementary Information). In the case of coherent combination, we obtain a trajectory that is sensitive to the relative phase between SBVs (Fig. 2c). In contrast, the combination of SOPs with greatly reduced mutual coherence, i.e. incoherent, traces a trajectory corresponding to the shortest path (geodesic) connecting the SOP of the initial to the final state on the Poincaré sphere, which is independent of relative phase (see Supplementary Information).

Coverage of the entire Poincaré sphere by SBVs comprised of four degenerate SOPs (the horizontal, vertical, +45°, and right circular polarizations) is shown in Fig. 2a,c. The regions enabled by each subset of three SBVs piece together to entirely cover the Poincaré sphere. However, SOP coverage (the angular change in SOP corresponding to a change in modulation parameters) is nonuniform for the set of degenerate SBVs (see Supplementary Material). We improved uniformity by borrowing from optimization techniques used in polarimetry[20–22]: optimal and minimal polarimetry and symmetric informationally complete positive operator valued measures (SIC-POVM). In these methods, a polarimeter measures the intensities of four states corresponding to the vertices of a regular tetrahedron inscribed in the Poincaré sphere. This arrangement maximizes the distance between measured states. When constructing a PSG with degenerate SBVs, the four SOPs define an irregular tetrahedron, resulting in a greater density of SOPs gathered around octant I of the Poincaré sphere. We calculated that a set of SBVs with elliptical SOPs defining a regular tetrahedron greatly improves uniformity of coverage compared with four degenerate SBVs (Fig. 2b,d).

To implement our method experimentally, a wide range of intensity modulators and wavelengths, as well as free-space, guided, and on-chip configurations are available to us. In our experiment, we used a digital micromirror device (DMD) to modulate four spatially separated SBVs derived from a laser beam to digitally generate a laser beam with arbitrary SOP (Fig. 3 and see Methods for details). We were able to generate coherent trajectories between SBVs (Fig. 4a). A Monte Carlo experiment was performed to probe coverage of SOPs over the Poincaré sphere with 200 random modulation parameters and produced good uniformity of coverage using a set of regular tetrahedral SBVs (Fig. 4b). A time-varying polarization signal was measured at slow speeds and matched well with the theory based on equation (3) (Fig. 4c). Measurements were also performed of the switching speed between linear horizontal and vertical SOPs, in which a high-speed pseudorandom bitstream was displayed on the DLP chip to generate an eye pattern (Fig. 4d and Supplementary Information).

Our main concern with the parallel architecture, yet, is insertion loss. Absorption or reflection modulators inherently use loss as a means of modulation. Additionally, coherent beam combining methods can only efficiently combine beams that are in-phase and have equal amplitude[23], and our architecture rarely combines beams that satisfy both requirements. However, improvements can be made easily to the modulation stage by using directional couplers[24] that retain all of the optical power when setting the relative modulation parameters between the SBVs. In the combination stage, a more sophisticated method is still sought to combine beams of varying amplitudes. Nonetheless, numerical calculations show that loss due to coherent beam combining is at a level that may be acceptable for applications in which the features of parallel polarization state generation are desirable. The average theoretical insertion loss by generating 80,000 SOPs distributed uniformly over the Poincaré sphere was calculated to be 6.5±4.4 dB for a set of 4 degenerate SBVs and 8.0±2.1 dB for a set of regular tetrahedral SBVs (see Supplementary Information).

In conclusion, we have introduced and experimentally implemented a parallel architecture for PSG, based on intensity modulation of separate polarization components. A major advantage is that the particular features of an embodiment are determined by the technology of intensity modulation used. For example, in our case, broadband metallic mirrors of the DMD used would translate to broadband PSG. Furthermore, figures of merit, such as speed and affordability, will continue to increase commensurately with modulator development: e.g., a system built with injection-locked directly modulated lasers[25]. It is interesting to note that the architecture can be inverted to form a conventional Stokes polarimeter, suggesting a polarization transceiver. In addition to foreseeing new applications in science and technology, analogous interference phenomena exist in quantum mechanics (as can be seen by the mathematical relationship of the Pauli matrices[26] and the coherency matrix[5] with the Stokes parameters, as well as the Bloch sphere with the Poincaré sphere), which may provide the potential to generalize this method to two-level quantum systems, such as coherent electronic and magnetic systems.

## Methods

In order to modulate the intensities of each of the four beams, a black and white image corresponding to a random binary matrix with an average value equal to the desired intensity modulation parameter was displayed on each quadrant of the DMD. The DMD was a Texas Instruments DLP3000. The displayed image was changed according to the desired SOP. The output was then measured using a free-space polarimeter (Thorlabs PAX5710).

Sources of error include vibration of optical components. The final polarization state is sensitive to the jitter in the relative phase between each of the four beams, and the average angular SOP error was measured to be 5.9° on the Poincaré sphere (Fig. 4a,c). The SOP profile along the interfering wavefront changes smoothly, due to slight misalignment between the four beams, causing the relative phase difference between the SBVs to vary slightly as a function of position. Vibration of the pinhole causes the output beam to be a sample of a changing portion of the preceding wavefront and leads to SOP error. Additionally, simultaneous sampling of multiple SOPs by the pinhole leads to multiple SOPs detected and integrated by the polarimeter, which decreases the degree of polarization, as can be seen with unpolarized light that is mathematically decomposed into two uncorrelated orthogonal elliptical SOPs[1].

The polarization-modulated beam was incident on a high-speed photodiode (Thorlabs DET100A) with a mounted linear polarizer, and the optical signal was measured on an oscilloscope (Agilent 54855A DSO) triggered by the automatic trigger signal of the DLP controller. Switching speed was measured up to the maximum speed allowed by the DLP3000 at 4 kHz without any degradation or impact on SOP signal quality. The measured settling time was extremely fast (3.5 µs), following an exponential for a 1 kHz bit stream, which reflects the settling time of the DMD. SOP noise was dominated by the instability of relative phase between interfering beams, which are best seen in the polarization trajectory measurements of Fig. 4a,c.

## References

1. Collett, E. *Field Guide to Polarization*. (Society of Photo Optical, 2005).
2. Goldstein, D. H. *Polarized Light, Third Edition*. (Taylor & Francis, 2010).
3. Clarke, D. Nomenclature of Polarized Light: Linear Polarization. *Appl. Opt.* **13**, 3–5 (1974).
4. Shurcliff, W. A. & Ballard, S. S. *Polarized light*. (Published for the Commission on College Physics by D. Van Nostrand, 1964).
5. Born, M. *et al. Principles of Optics: Electromagnetic Theory of Propagation, Interference and Diffraction of Light*. (Cambridge University Press, 2000).
6. Imai, T., Nosu, K. & Yamaguchi, H. Optical polarisation control utilising an optical heterodyne detection

scheme. *Electron. Lett.* **21,** 52–53 (1985).

7.  Holmes, D. Wave Optics Theory of Rotary Compensators. *J. Opt. Soc. Am.* **54,** 1340 (1964).

8.  Lefevre, H. C. Single-mode fibre fractional wave devices and polarisation controllers. *Electron. Lett.* **16,** 778–780 (1980).

9.  Okoshi, T., Cheng, Y. H. & Kikuchi, K. New polarisation-control scheme for optical heterodyne receiver using two Faraday rotators. *Electron. Lett.* **21,** 787–788 (1985).

10. Ulrich, R. Polarization stabilization on single-mode fiber. *Appl. Phys. Lett.* **35,** 840–842 (1979).

11. Takasaki, H. & Yoshino, Y. Polarization interferometer. *Appl. Opt.* **8,** 2344–2345 (1969).

12. Lizana, A. *et al.* Arbitrary state of polarization with customized degree of polarization generator. *Opt. Lett.* **40,** 3790 (2015).

13. Kubota, M., Oohara, T., Furuya, K. & Suematsu, Y. Electro-optical polarisation control on single-mode optical fibres. *Electron. Lett.* **16,** 573 (1980).

14. Zhuang, Z., Suh, S.-W. & Patel, J. S. Polarization controller using nematic liquid crystals. *Opt. Lett.* **24,** 694–696 (1999).

15. Dong, P., Chen, Y.-K., Duan, G.-H. & Neilson, D. T. Silicon photonic devices and integrated circuits. *Nanophotonics* **3,** 215–228 (2014).

16. Miller, D. a. B. Self-configuring universal linear optical component. *Photonics Res.* **1,** 1 (2013).

17. Okoshi, T. Polarization-state control schemes for heterodyne or homodyne optical fiber communications. *J. Light. Technol.* **3,** 1232–1237 (1985).

18. Fresnel, A. J., de Sénarmont, H. H., Verdet, É. & Fresnel, L. F. *Œuvres complètes d'Augustin Fresnel: Théorie de la lumière.* (Imprimerie impériale, 1866).

19. Collett, E. Mathematical Formulation of the Interference Laws of Fresnel and Arago. *Am. J. Phys.* **39,** 1483 (1971).

20. Azzam, R. M. a., Elminyawi, I. M. & El-Saba, a. M. General analysis and optimization of the four-detector photopolarimeter. *J. Opt. Soc. Am. A* **5,** 681–689 (1988).

21. Sabatke, D. S. *et al.* Optimization of retardance for a complete Stokes polarimeter. *Opt. Lett.* **25,** 802–804 (2000).

22. Renes, J. M., Blume-Kohout, R., Scott, a. J. & Caves, C. M. Symmetric informationally complete quantum measurements. *J. Math. Phys.* **45,** 2171–2180 (2004).

23. Fan, T. Y. The Effect of Amplitude (Power) Variations on Beam Combining Efficiency for Phased Arrays. *IEEE J. Sel. Top. Quantum Electron.* **15,** 291–293 (2009).

24. Lu, Z. *et al.* Broadband silicon photonic directional coupler using asymmetric-waveguide based phase control. *Opt. Express* **23,** 3795 (2015).

25. Liu, Z. *et al.* Modulator-free quadrature amplitude modulation signal synthesis. *Nat. Commun.* **5,** 5911 (2014).

26. Fano, U. A stokes-parameter technique for the treatment of polarization in quantum mechanics. *Phys. Rev.* **93,** 121–123 (1954).

**Acknowledgements**

We thank S. Zhang, J. Chee, J. Mueller, J.P. Laine, A. Cable, and C.-E. Zah for discussions. The authors thank D. Ham for providing the Agilent 54855A DSO oscilloscope. A.S. is supported by the Harvard School of Engineering & Applied Sciences Winokur Fellowship (2012-2014) and the Charles Stark Draper Laboratory through the Draper Lab Fellowship (2014-present). This work was partially supported by the Intelligence Advanced Research Projects Activity (IARPA) under grant N66001-12-C-2011.

**Author contributions**

Alan She and Federico Capasso

**Competing financial interests**

The authors declare no competing financial interests.

## Figures

**(a)**

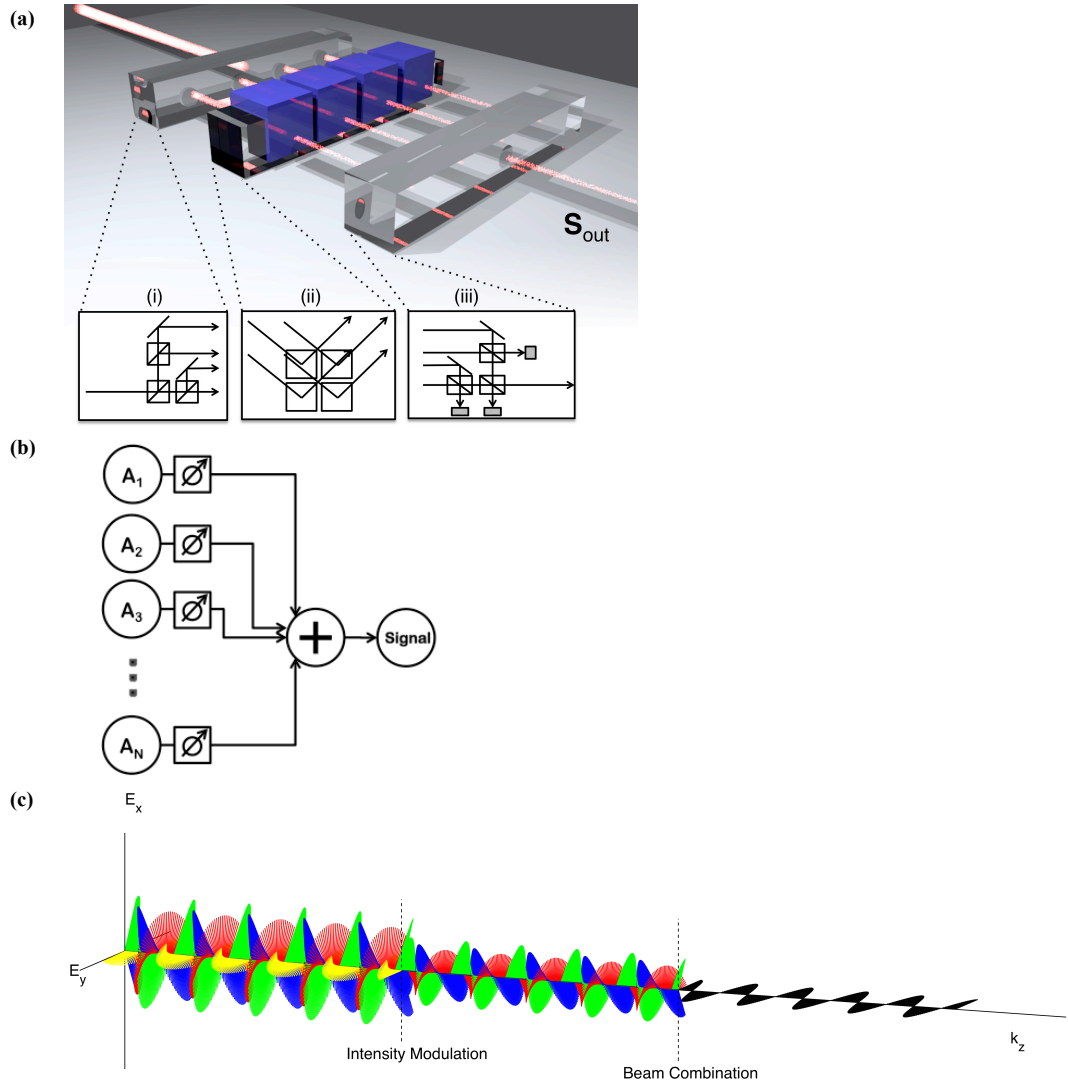

**Figure 1 | Concept. (a)** An illustration showing the general, modular implementation of the described method for a parallel polarization state generator (PSG). An input beam is (i) split into four beams of different polarizations, which are then (ii) intensity modulated either in reflection or transmission, (iii) and finally combined to form a single output beam, the polarization and phase of which can be tuned with a precision and speed limited by the modulator. **(b)** A schematic of PSG architecture is shown, in which modulators are placed after light sources $A_i$ with well-defined states of polarization (SOP) and relative phase, and their weighted linear superposition produces the desired output signal. **(c)** Generation of horizontally polarized light using this method is illustrated. The electric fields of four propagating electromagnetic waves (red, green, blue, and yellow) with elliptical polarizations are superimposed and plotted as function of wave propagation position. They are intensity modulated and beam combined to generate the desired horizontal polarization (black).

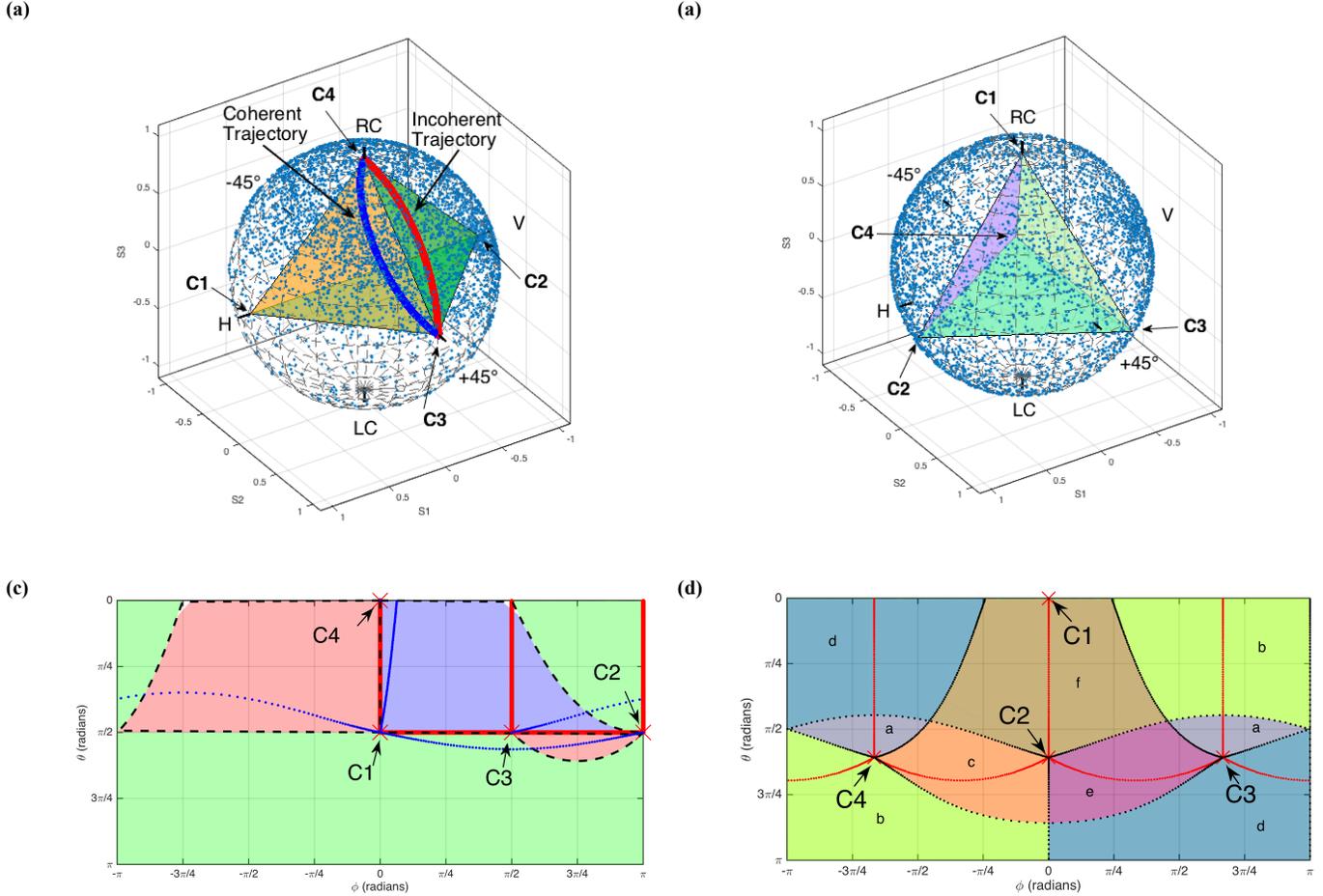

**Figure 2 | Simulations.** Two systems with distinct sets of Stokes basis vectors (SBVs) were simulated– one composed of degenerate SOPs and the other with SOPs mapped to a regular tetrahedron on the Poincaré sphere. Intensity modulation parameters of individual SBVs are varied to generate polarization trajectories and whole regions of accessible SOPs. (**a**) A system with SBVs with four degenerate SOPs: linear horizontal (C1), vertical (C2), +45° with a 180° phase shift (C3), and right circular polarization (C4), are shown. A Monte Carlo simulation (blue points) was performed by randomly varying the intensity modulation parameters and showed complete, yet non-uniform coverage of SOPs over the Poincaré sphere. A polarization trajectory between SBVs C3 to C4 is shown for coherent combination (blue line) and incoherent combination (red line). Incoherent trajectories are geodesics. (**b**) A system with SBVs optimized for better uniformity of SOP coverage is shown, in which the SBVs are vertices of a regular tetrahedron inscribed in the Poincaré sphere, as opposed to the degenerate SBVs. In Jones vector notation, the SBVs used here were [0.7071, 0.7071i], [-9.856, 0.1691i], [0.5141, 0.7941-0.3242i], and [0.5141, -0.7941-0.3242i], labeled as C1-4, respectively. (**c**) The degenerate system of (a) is mapped using a Mercator projection of the Poincaré sphere, where $\theta$ is the polar angle and $\phi$ is the azimuthal angle. The locations of the SBVs here, C1-4, are the same as in (a). All coherent and incoherent polarization trajectories between SBVs are shown with black dotted and red solid lines, respectively. The coherent polarization trajectories connected to C1 are warped by increasing the relative phase difference between C1 and other SBVs by 6° (blue dotted lines). The colored regions show the regions of SOPs enabled by combinations of three SBVs: by combining C1, C2, and C4, with varying intensity modulation parameters, all SOPs in the blue region can be generated; similarly, combinations of (C1, C3, C4) and (C2, C3, C4) generate the red and green regions, respectively. However, (C1, C2, C3) generate a region of no area because these SBVs are not linearly independent in this particular system. (**d**) The regular tetrahedron system of (b) is mapped using a Mercator projection of the Poincaré sphere. Coherent and incoherent polarization trajectories between SBVs are shown with black and red dotted lines, respectively. In this case, regions of SOPs generated by combinations of sets of three SBVs are well distributed and have similar size and great overlap, yielding better overall uniformity. Due to great overlap between regions, they are color coded and labeled as the following: C1, C2, C3 combine to cover regions a, b, and c; similarly: C1, C2, C4 (a, d, e); C1, C3, C4 (c, e, f); and C2, C3, C4 (b, d, f).

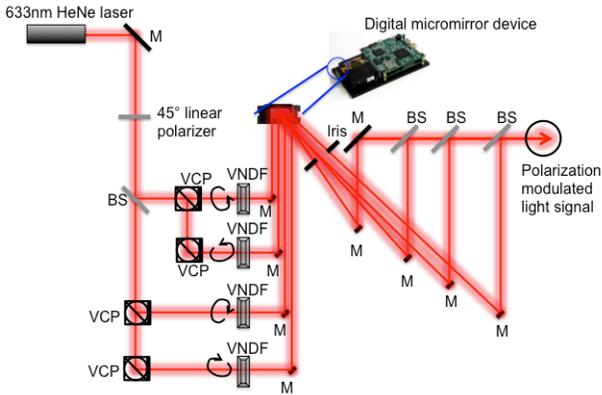

**Figure 3 | Experimental setup.** Light from a HeNe laser is prepared in the linear +45° polarization using a wire-grid polarizer. The beam is then split into two beams by a non-polarizing beam splitter (BS). Each of these beams is split again using variable circular polarizers (VCPs) into two intensity-modulated polarization states. The resultant SOP of the four beams is tuned by rotating the quarter wave plate embedded in the VCPs. Variable neutral-density filters (VNDFs) are placed directly after the VCPs to balance the four beam intensities. The four beams are then directed onto four quadrants of the surface of a computer controlled Texas Instruments DLP3000 digital micromirror device (DMD). The DMD is composed of an array of polarization-insensitive mirrors that can be switched in one of two positions. Mirrors that point in the direction of the output beam contribute to the total intensity and all other light is directed into a beam dump. The DMD behaves as a 2-D diffraction grating for the incident laser light. An iris is used to select the strongest diffraction order. The path length differences of the four intensity-modulated beams passing through the iris are adjusted to be less than the coherence length of the laser (< 20 cm) with a series of mirrors. They are combined using three non-polarizing beam splitters to form a single beam. Finally, this beam is passed through a 100-μm pinhole, in order to select a small uniform portion of the wavefront of the combined beam to maximize the degree of polarization, to form the PSG output.

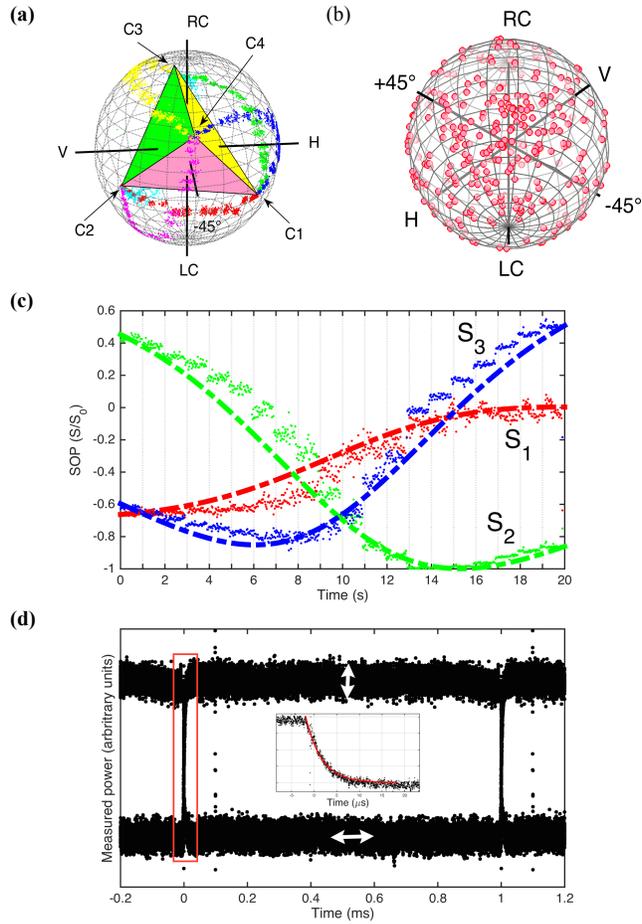

**Figure 4 | Experimental results. (a)** Data from the experimental setup of Fig. 3. The Stokes basis vectors (SBVs) are set to SOPs approximating (within the error of tuning the variable circular polarizers) a regular tetrahedron on the Poincaré sphere. The SBVs C1, C2, C3, and C4 were measured and the resulting tetrahedron is drawn. Coherent polarization trajectories from each SBV to every other SBV are generated by modulating SBV intensities in 20 discrete increments spanning 20 seconds, and the raw data as measured by the polarimeter are shown. **(b)** The results of a Monte Carlo experiment, in which 200 random intensity modulation parameters $\alpha$ were used, are shown on the Poincaré sphere, indicating good uniformity of coverage of SOPs. **(c)** Time series data of a coherent polarization trajectory between two SBVs (C2 to C4) in (a) are compared to theoretical calculation (dotted line) and show good agreement. $S_1$, $S_2$, and $S_3$ are elements of the Stokes vector. **(d)** An eye pattern is generated for a polarization signal that switches between linear horizontal and vertical polarizations using the DLP3000. The data are shown for a pseudorandom bitstream modulated at 1 kHz. The inset is a larger view of the red rectangle and shows the measured settling time (eye rise and fall time) to be 3.5 μs following an exponential.

Supplementary Information

**Parallel Polarization State Generation**

Alan She[1]* and Federico Capasso[1]*

[1]Harvard John A. Paulson School of Engineering and Applied Sciences, Cambridge, MA 02138

*To whom correspondence should be addressed;
E-mail: ashe@seas.harvard.edu; capasso@seas.harvard.edu.

## S1    Polarization mathematics

Polarization is commonly represented in two forms: the Jones vector and the Stokes parameters. The Jones vector is a complex 2-element vector, which describes completely polarized light by defining the phases and amplitudes of two orthogonal electric field components. The electrical field of a plane wave propagating in the z direction can be written as the following:

$$\mathbf{E}(z,t) = \left( E_{0x} \cos(kz - \omega t + \delta_x), E_{0y} \cos(kz - \omega t + \delta_y) \right) \tag{S1}$$

The complex amplitude coefficients of equation (S1) are known as the Jones vector: $\left( E_{0x} e^{i\delta_x}, E_{0y} e^{i\delta_y} \right)$. The Stokes parameters, which were defined by G. Stokes in 1852 to mathematically describe polarized light, including partial polarization[1], are extremely useful today but were historically hampered by the inability to quantify optical intensity measurements. To derive the Stokes parameters, the equation of the polarization ellipse[2], $\frac{E_x^2}{E_{0x}^2} + \frac{E_y^2}{E_{0y}^2} - 2 \frac{E_x}{E_{0x}} \frac{E_y}{E_{0y}} \cos\delta = \sin^2\delta$, where $\delta = \delta_y - \delta_x$, can be rearranged such that the grouped terms of $(E_{0x}^2 + E_{0y}^2)^2 - (E_{0x}^2 - E_{0y}^2)^2 - (2E_{0x}E_{0y}\cos\delta)^2 = (2E_{0x}E_{0y}\sin\delta)^2$ can be written as the following Stokes parameters:

$$\begin{aligned} S_0 &= E_{0x}^2 + E_{0y}^2 \text{ ,} \\ S_1 &= E_{0x}^2 - E_{0y}^2 \text{ ,} \\ S_2 &= 2E_{0x}E_{0y}\cos\delta \text{ , and} \\ S_3 &= 2E_{0x}E_{0y}\sin\delta \text{ .} \end{aligned} \tag{S2}$$

$S_0$ is the total intensity, and $\left( S_1, S_2, S_3 \right)$ is the Stokes vector describing the SOP. The Stokes parameters simplify measurement of SOP enormously by requiring only 4 intensity measurements. A triangle inequality exists, in which the total intensity $S_0^2 \geq S_1^2 + S_2^2 + S_3^2$. The ratio of the length of the Stokes vector to the total intensity is the degree of polarization: $DOP = \frac{\sqrt{S_1^2 + S_2^2 + S_3^2}}{S_0}$. The Stokes vector that is normalized to a unit vector traces all possible SOPs on a mathematical object called the Poincaré sphere.

## S2    Regions of coverage of Poincaré Sphere using degenerate SBVs

The degenerate SBVs take on any four of the six degenerate polarizations, which are the linear horizontal, linear vertical, linear +45°, linear -45°, right circular, and left circulation polarizations. We explored a system of SBVs in the four following polarizations: linear horizontal, linear vertical, linear +45°, and right circular. Coverage of the Poincaré sphere by possible SOP states using the above system is shown in Fig. 2a,c. Fig. S1 plots the Mercator projection of Fig. 2c on the Poincaré sphere. Each system is uniquely defined by the SOPs and the global phases $\phi_n$ of the SBVs that comprise it. In this particular system, all SBVs have $\phi=0°$, except the SBV with linear +45° SOP, for which $\phi=180°$.

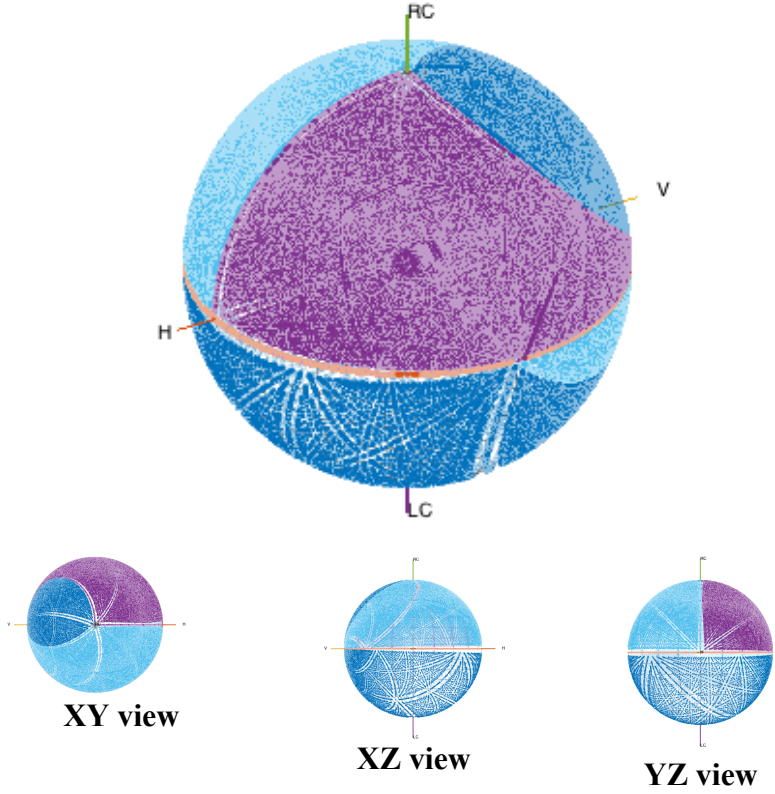

Figure S1: **Degenerate SBV coverage.** The Poincaré sphere is shown, covered by possible SOPs, as generated by linear combinations of four degenerate SBVs, in the following polarizations: linear horizontal, linear vertical, linear +45°, and right circular. All SBVs have global phases $\phi$=0°, except that of the linear +45° polarization with $\phi$=180°. This system is identical to that of Fig. 2a,c.

## S3     Discussion of metric tensor

The uniformity of the system can be described by the metric on the Poincaré sphere, where angular separation between SOPs in Stokes space is given by $\cos\theta_{AB} = \dfrac{\mathbf{S}_A \cdot \mathbf{S}_B}{|\mathbf{S}_A||\mathbf{S}_B|}$ and in Jones vector space $\cos^2\dfrac{\theta_{AB}}{2} = \dfrac{\left|\langle E_A | E_B \rangle\right|^2}{\langle E_A | E_A \rangle \langle E_B | E_B \rangle}$. By describing nearby states as $E_i = C_{ij}\alpha_j$ and $E_i = C_{ij}\left(\alpha_j + \delta\alpha_j\right)$, it is possible to construct a metric tensor, where $ds^2 = d\theta^2$. For the case of coherent combination, this is the Fubini-Study metric, and for incoherent combination this is the Bures metric.

## S4     Coherent versus incoherent combination

As described in the main text, coherent combination (Fig. S2) produces polarization trajectories that are sensitive to the difference in global phases $\phi_n$ between SBVs (Fig. S3), whereas incoherent combination necessitates a geodesic trajectory that is insensitive to $\phi_n$ (Fig. S4). It is possible to switch between these two combination methods to generate trajectories with degrees of coherence intermediate between the two limits by changing the mutual coherence between SBVs. Mutual coherence, or the degree of cross-correlation, can be tuned by varying the optical path length between SBVs. By having the optical length exceed the coherence length of the light source, the SBVs no longer have a fixed phase relation and incoherently combine. Fig. S4 further illustrates the difference in polarization trajectories as generated by the two methods of combination.

The SOP measured following incoherent combination can be described as a linear sum of Stokes vectors: $\mathbf{S}_{out} = \alpha_1^2 \mathbf{S}(\mathbf{C}_1) + \alpha_2^2 \mathbf{S}(\mathbf{C}_2) + \alpha_3^2 \mathbf{S}(\mathbf{C}_3) + \alpha_4^2 \mathbf{S}(\mathbf{C}_4)$, in which $\mathbf{S}(\mathbf{C}_1)$ is the Stokes vector corresponding to the SOP of $\mathbf{C}_1$, and so on. This can be seen as the simultaneous detection of the intensities of non-interfering beams with their unaltered SOPs; hence the intensities of Stokes vectors add linearly on the detector side. The incoherent trajectory is insensitive to the relative phase difference between SBVs. Finally, the degree of polarization of the generated SOP as measured by a polarimeter has two contributors: (a) any unpolarized background originating from the unpolarized parts of the sources' signals and (b) any less than unity value of the degree of cross-correlation (mutual coherence) between combined SBVs.

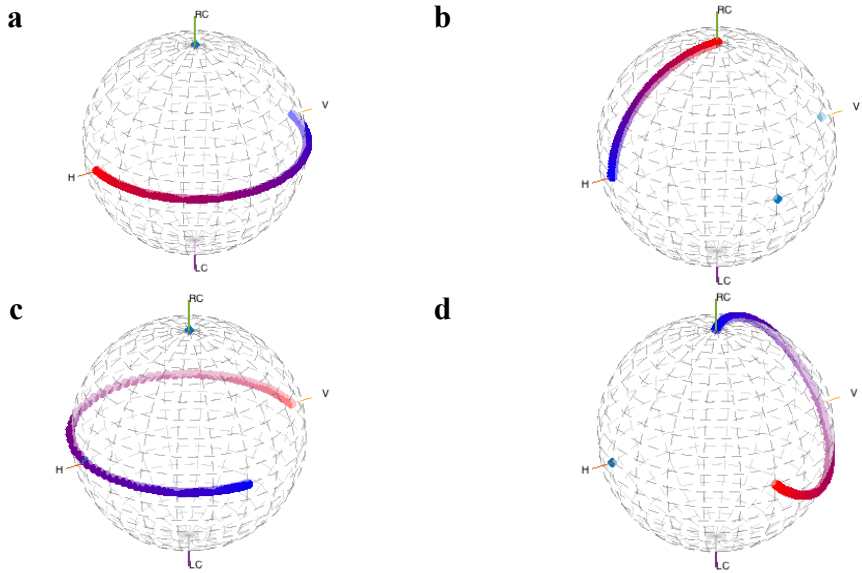

Figure S2: **Coherent polarization trajectories on the Poincaré sphere.** Trajectories are shown, where the polarization is varied from the initial state (red) to the final state (blue). Four trajectories between degenerate SOPs are plotted: **a)** linear horizontal to linear vertical, **b)** right circular to linear horizontal, **c)** linear vertical to linear +45°, and **d)** linear +45° to right circular. The global phase of the linear +45° polarization is defined to be 180° out-of-phase with respect to the other degenerate SOPs, such that in Jones vector notation it is [-1, -1].

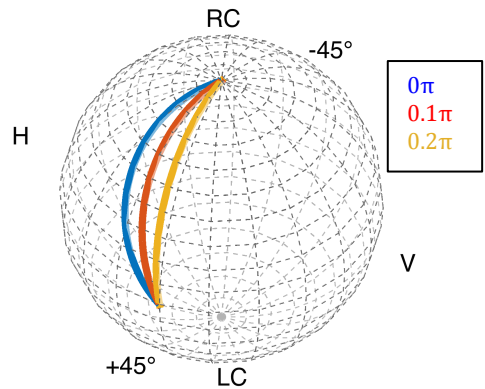

Figure S3: **Phase dependence of coherent polarization trajectories on the Poincaré**

**sphere.** The effect on polarization trajectories by changing the relative phase between two linearly combined SOPs is shown. The polarization trajectory connecting SOPs can be modified in either direction by controlling the relative phase.

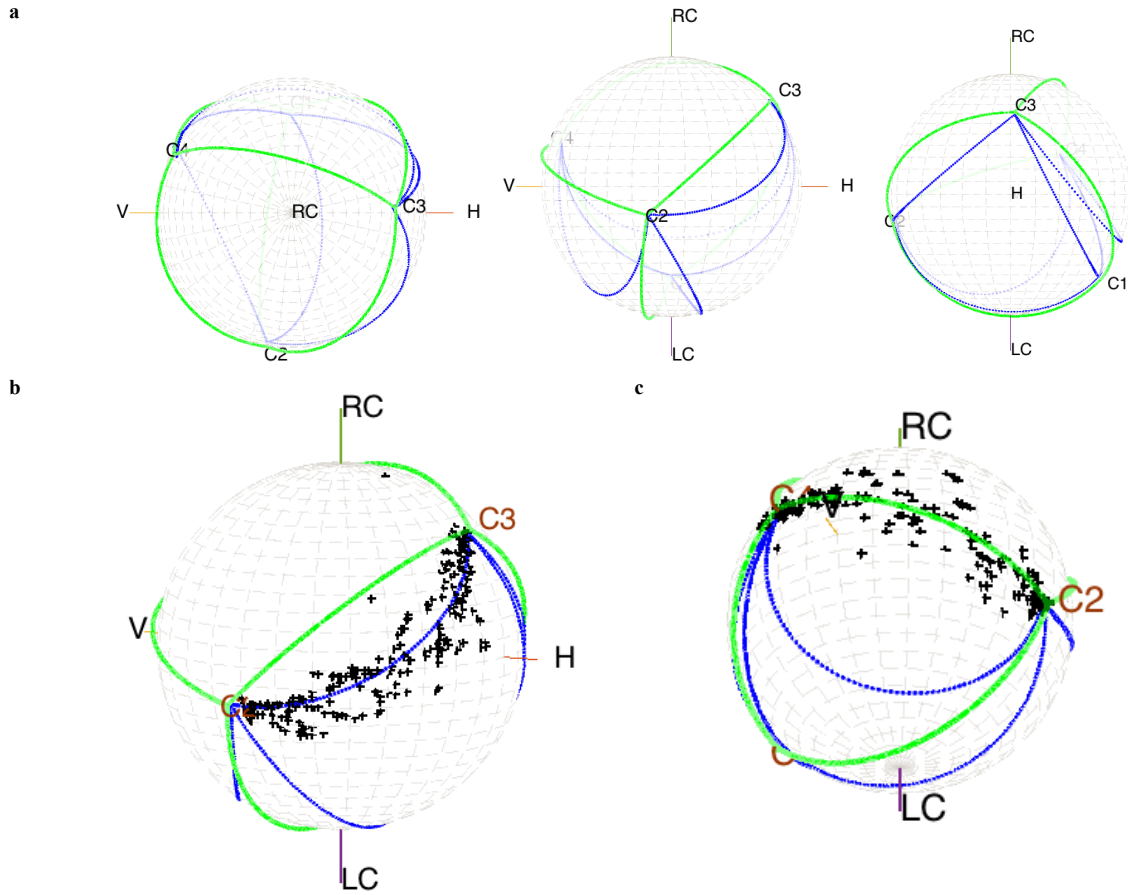

Figure S4: **Coherent and incoherent polarization trajectories on the Poincaré. a)** Various perspective views of an example system. The SBVs used are labeled C1-4. Polarization trajectories are generated by coherent combination (blue line) and incoherent combination. **b)** This example system is implemented using the experimental setup described in the main text. A polarization trajectory is generated by coherent combination by keeping the optical path length between the two SBVs C2 and C3 well below the coherence length of the laser (~20 cm). The trajectory is measured by the polarimeter and the data are shown. **c)** An incoherent polarization trajectory following the geodesic path between SBVs C2 and C4 is generated by making the optical path length between SBVs much longer than the coherence length of the laser (hence reducing the mutual coherence).

## S5    Comparison of the performance of our experiment with a commercial PSG

Performance characteristics of our implementation are promising, with an SOP settling time (representing speed and stability) of 3.5 μs compared to a state-of-the-art device (Thorlabs DPC5500) with 150 μs for < 10° deviation and 1 ms for < 1° deviation. However the SOP accuracy of our embodiment (5.9° error) is limited by the unstable relative phase between the four SBVs, whereas the DPC5500 can be as accurate as 0.25°. However, there is room for major improvements, in terms of both speed and accuracy, such as by using faster modulators and miniaturization; the latter would greatly increase the phase stability between SBVs and reduce the error. Realistically, we expect PSGs in the visible and telecom wavelengths, for example, to achieve the speeds of the fastest modulators available, e.g. greater than 40 GHz (lithium niobate), pushing PSG technology from the kiloradians/second regime into the gigaradians/second.

**S6      Insertion loss calculation**

The fundamental limitation to the efficiency of the parallel architecture stems from inefficient beam combining. The theoretical coherent beam combining efficiency for a system with N ports has been derived as the following[3]:

$$\eta = \frac{1}{N} \frac{\left| \sum_{m=1}^{N} \sqrt{P_m} \exp(j\phi_m) \right|^2}{\sum_{m=1}^{N} P_m} \tag{S3}$$

where $P_m$ is the power and $\phi_m$ the phase of the $m$th beam. This assumes perfect alignment and coherence, so, in practice, there will be additional losses. Theoretical insertion loss ($IL$) was calculated using equation (S3), where $IL = -10 \log(\eta)$, for a large number of SOPs which represented uniform coverage of the Poincaré sphere. Descriptive statistics are shown in Table S1.

A PSG constructed from a set of 4 degenerate SBVs has a greater range of insertion loss compared to that of regular tetrahedral SBVs (Fig. S5 and see standard deviation values in Table S1). This can be explained by the increased uniformity of the regular tetrahedral SBVs, which are well-separated SOPs in state space, in contrast to the degenerate SBVs, implying that on average each generable SOP will be much closer to one single SBV than the other three, leading to a large yet frequent power imbalance between the beams when combining. With the set of 4 degenerate SBVs, there is greater variability in the state space distance from each generable SOP to the SBVs.

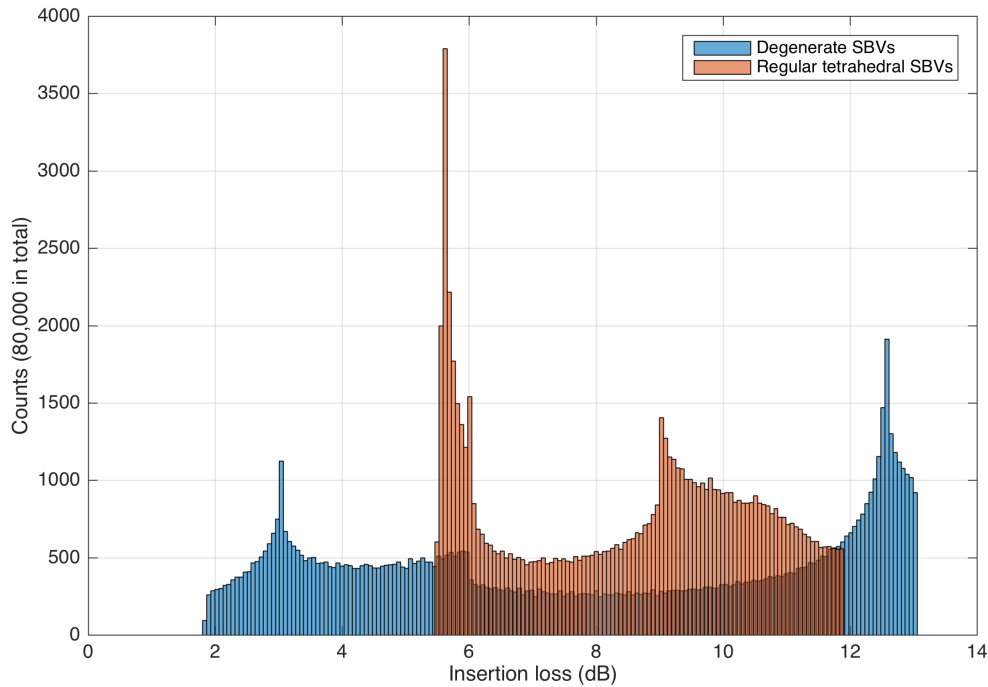

Figure S5: **Insertion loss calculation.** Insertion loss is calculated by solving equation (3) and equation (S3) for 80,000 SOPs distributed uniformly over the Poincaré sphere. The distribution of insertion losses are represented by histograms, for which a set of 4 degenerate SBVs and a set of regular tetrahedral SBVs are shown in blue and orange, respectively.

Table S1: **Insertion loss statistics for Fig. S5.**

| | Degenerate SBVs | | Regular tetrahedral SBVs | |
|---|---|---|---|---|
| | Efficiency (linear) | Insertion loss (dB) | Efficiency (linear) | Insertion loss (dB) |
| Mean | 0.22 | 6.50 | 0.16 | 8.04 |
| Minimum | 0.05 | 13.05 | 0.06 | 11.89 |
| Maximum | 0.65 | 1.84 | 0.28 | 5.47 |
| Standard deviation | 0.17 | 4.47 | 0.07 | 2.11 |

**References**

1.  Stokes, G. G. On the composition and resolution of streams of polarized light from different sources. *Trans. Cambridge Philos. Soc.* **IX,** 233–58 (1852).
2.  Collett, E. *Field Guide to Polarization.* (Society of Photo Optical, 2005).
3.  Link, C., Fan, T. Y. & Member, S. The Effect of Amplitude ( Power ) Variations on Beam Combining Efficiency for Phased Arrays. (2016). doi:10.1109/JSTQE.2008.2010232